%% file: conference_101719.tex
\documentclass[conference]{IEEEtran}
\IEEEoverridecommandlockouts
\usepackage{cite}
\usepackage{amsmath,amssymb,amsfonts}
\usepackage{algorithmic}
\usepackage{graphicx}
\usepackage{textcomp}
\usepackage{xcolor}

\usepackage{tikz}
\usepackage{longtable}
\usepackage{booktabs}
\usepackage{array}
\usepackage{epstopdf}
\usepackage[utf8]{inputenc}

\usepackage{todonotes}

\usepackage[norelsize, ruled,vlined]{algorithm2e}
\SetKwProg{Init}{Initialization:}{}{}

\DeclareMathOperator*{\tw}{tw}

\newcommand{\edp}{{\sc{Edge Disjoint Paths}}\xspace}
\newcommand{\ndp}{{\sc{Node Disjoint Paths}}\xspace}

\newcommand{\duspr}{{\sc{\mbox{D-USPR}}}\xspace}
\newcommand{\kduspr}{ \mbox{\sc{Pre Routed D-USPR} }\xspace}

\newlength{\atextwidth}
\newcommand{\rc}{\ensuremath{H}}
\newcommand{\rcset}{\ensuremath{\mathcal{H}}}

\newcommand{\dc}{\ensuremath{d}}
\newcommand{\dcset}{\ensuremath{\mathcal{D}}}

\newcommand{\problemopt}[3]{
  \vspace{1mm}
\noindent
  \begin{minipage}{0.47\atextwidth}
  #1 \\
  {\bf{Input:}} #2  \\
  {\bf{Output:}} #3
  \end{minipage}
  
  \vspace{1mm}
}

\usetikzlibrary{chains,positioning,backgrounds,shapes,fit}

\tikzstyle{background}=[rectangle,fill=red!10,inner sep=0.2cm,rounded corners=1mm,minimum size=1cm]

\tikzstyle{vertex}=[circle,thick,draw,minimum size=5pt,inner sep=0pt,fill=black]
\tikzstyle{terminal}=[vertex,rectangle]
\tikzstyle{selvertex2}=[rectangle,rounded corners=3mm,draw=darkred,inner sep=5pt, ultra thick]
\tikzstyle{selected vertex} = [vertex, ultra thick]
\tikzstyle{activated vertex} = [vertex, fill=red!50]
\tikzstyle{edge} = [draw,ultra thick,-]
\tikzstyle{route} = [draw,thick,-,line width=0.3mm, draw=black!50]
\tikzstyle{tableedge} = [draw,thick,-,dashed,black!75]
\tikzstyle{weight} = [font=\small]
\tikzstyle{selected edge} = [draw,line width=5pt,-,red!50]
\tikzstyle{ignored edge} = [draw,line width=5pt,-,black!20]
\tikzstyle{tmonnode} = [draw,ultra thick,minimum size=7mm]
\tikzstyle{tmnode} = [draw,minimum size=6.5mm]
\tikzstyle{bag}=[rectangle,rounded corners=3mm,draw=black,inner sep=5pt, thick]

\newtheorem{definition}{Definition}

\newtheorem{proposition}{Proposition}
\newtheorem{claim}{Claim}
\newtheorem{lemma}{Lemma}

\newcommand{\dynalgotimek}{$2^{|K|\omega^8 + \Delta \log |K|} \cdot n^{O(1)}$}

\def\BibTeX{{\rm B\kern-.05em{\sc i\kern-.025em b}\kern-.08em
    T\kern-.1667em\lower.7ex\hbox{E}\kern-.125emX}}
\begin{document}

\title{Toward Scalable Algorithms for the Unsplittable Shortest Path Routing Problem\\
}

\author{\IEEEauthorblockN{Amal Benhamiche}
\IEEEauthorblockA{\textit{Data $\&$ Artificial Intelligence,} \\
\textit{Orange Labs,}\\
Ch\^{a}tillon, France, \\
amal.benhamiche@orange.com}
\and
\IEEEauthorblockN{Morgan Chopin}
\IEEEauthorblockA{\textit{Data $\&$ Artificial Intelligence,} \\
\textit{Orange Labs,}\\
Ch\^{a}tillon, France, \\
morgan.chopin@orange.com}
}

%Orange Labs, Ch\^{a}tillon, France,\\
%\email{\{amal.benhamiche,morgan.chopin\}@orange.com},\\
%Orange Gardens, 44 avenue de la R\'{e}publique 92326 Ch\^{a}tillon Cedex France

\maketitle

%First, the objective function need not to be linear, and this allows, for instance, to tackle natural multicriteria versions of the problem. For instance, it would be possible to use non-linear aggregator to compute routing paths that are balanced with respect to several criterion such as the congestion, the delay and the costs of using some edges. 
%Second, this algorithm avoids to recompute from scratch a new set of routing paths in the case of a failure scenario \textsl{i.e.} a link in the network is down. Indeed, the tree decomposition makes it easy to identify what parts of the calculation need to be updated.
%Finally, even if the running time seems prohibitive, this algorithm can be used to route only a limited amount of demands. For instance, those that induce the highest congestion in the edges. Doing so, we reduce the impact of~$|D|$ in the running time exponent and if applied on tree-like networks, the algorithm scales polynomially with respect to the number of nodes.

\begin{abstract}
In this paper, we consider the \textit{Delay Constrained Unsplittable Shortest Path Routing} problem which arises in the field of \textit{traffic engineering} for IP networks. This problem consists, given a directed graph and a set of commodities, to compute a set of routing paths and the associated administrative weights such that each commodity is routed along the unique shortest path between its origin and its destination, according to these weights. We present a compact MILP formulation for the problem, extending the work in \cite{Bley2010} along with some valid inequalities to strengthen its linear relaxation. This formulation is used as the bulding block of an iterative approach that we develop to tackle large scale instances. We further propose a dynamic programming algorithm based on a \textit{tree decomposition} of the graph. To the best of our knowledge, this is the first exact combinatorial algorithm for the problem. Finally, we assess the efficiency of our approaches through a set of experiments on state-of-the-art instances.     
\end{abstract}

\section{Introduction}
\label{sec:intro}
\input{intro.tex}

\section{Preliminaries}
\label{sec:preliminaries}
\input{preliminaries.tex}

\section{Description of the problem}
\label{sec:problem_description}
\input{problem_description.tex}

\section{MILP formulation}
\label{sec:milp_formulation}
\input{milp_formulation.tex}

\section{An effective iterative algorithm}
\label{sec:iterative_algorithm}
\input{iterative_algorithm.tex}

\section{A dynamic programming algorithm}
\label{sec:dynamic_programming_algo}
\input{dynamic_prog.tex}

\section{Numerical results}
\label{sec:numerical_results}
\input{numerical_results.tex}

\section{Concluding remarks}
\label{sec:conclusion}
\input{conclusion.tex}

\bibliographystyle{abbrv}

\end{document}

%% file: intro.tex
In spite of the promises of the MPLS\footnote{\textit{MultiProtocol Label Switching}} forwarding scheme, most IP networks still heavily rely on shortest path rules where weights are assigned to links by network administrators and the routers are then able to compute shortest routing paths \cite{Perrot2019}. At the same time, the growing interest for content and user services driving even more traffic stresses the need to optimize the utilization of the available resources and maintain a high level of QoS on operational networks. On another hand, the arising of Network Virtualization will be a key enabler for the deployment of virtualized components capable of performing efficient path computation on behalf of the routers, thus allowing the optimization of operational IP networks. This perspective change  draws again the traffic engineering community's attention to classical problems related to IP network optimization and raises the question of finding effective algorithms allowing to solve those problems for large scale networks.

The problem of finding routing weights inducing shortest paths that minimize the network congestion has been widely studied in the literature for both its \textit{splittable} and \textit{unsplittable} versions. Some early works addressing shortest path routing issues in an IP network optimization include \cite{Bley1998}. The authors of this paper study the problem of designing a survivable VPN-based network using OSPF\footnote{Open Shorest Path First} routing protocol and propose a compact MILP formulation and several heuristics to solve the problem. In \cite{Fortz2000}, the authors investigate the OSPF weights optimization problem with splittable traffic and a piecewise approximation of the load function. They show that the problem is NP-hard for a given set of demands and provide a local search heuristic to solve it. The problem is also shown NP-hard to approximate for both splittable and unsplittable versions (see \cite{Fortz2000} and \cite{bley2009}). The authors in \cite{benameur03} and \cite{BleyPhdthesis} adress the unsplittable shortest path routing problem and study the properties of path sets that induce shortest path routing with compatible weights. Approaches based on Mixed Integer Linear programming for the problem include the work in \cite{Parmar2006}, \cite{Bley2010}, \cite{Bley2011}. In particular, compact formulations are proposed in \cite{Parmar2006} and \cite{Bley2010} for the splittable version of the problem. Bley propose in \cite{Bley2010} the so-called \textit{two-phase algorithm}, an exact approach based on a decomposition of the problem into a master subproblem and a client subproblem. The former subproblem generates routing paths while the latter returns compatible weights if any, or \textit{conflict inequalities} forbidding incompatible routing paths otherwise. \\

In this paper, we consider a variant of the \textit{Unsplittable Shortest Path Routing} (USPR) problem with end-to-end delay constraints motivated by practical QoS requirements for the traffic. Our work extends the results proposed by \cite{Bley2010} and \cite{Bley2011}. Our main contributions are $(i)$ a compact MILP formulation for the problem along with two classes of valid inequalities to strengthen its linear relaxation, $(ii)$ a MILP-based heuristic that iteratively reroutes a portion of the traffic fixed by the decision-maker and reduces the network load and $(iii)$ a dynamic programming algorithm based on a tree decomposition of the graph. To the best of our knowledge, this is the first exact combinatorial algorithm for the problem.\\

Our algorithms are designed with the two following objectives
\begin{itemize}
    \item[$\bullet$] \textbf{scalability}: they can be used to push back the limits of existing approaches for the problem in terms of size of instances treated,
    \item[$\bullet$] \textbf{flexibility}: they can be parameterized to provide solutions that modify only part of the traffic routing, which is highly desirable in practice. 
\end{itemize}
In addition, they can be either integrated in a centralized entity capable of computing intra-domain routing strategies that optimize the network load like a Path Computation Element (PCE) or SDN controller, or used as an off-line tool by the decision-makers for network planning operations. \\

The remainder of the paper is organized as follows. Section \ref{sec:preliminaries} is devoted to preliminaries and basic definitions. In Section \ref{sec:problem_description}, we introduce some necessary notations and give a formal definition of the problem along with a compact MILP formulation. We further present two families of inequalities valid for the problem. We describe our MILP-based iterative solving approach in Section \ref{sec:iterative_algorithm} and a dynamic programming algorithm in Section \ref{sec:dynamic_programming_algo}. Finally, Section \ref{sec:numerical_results} is devoted to present some early experiments to show the efficiency of our algorithms for state-of-the-art instances while some concluding remarks are given in Section \ref{sec:conclusion}.

%% file: preliminaries.tex
In this section, we give the graph notations and notions used throughout this paper

\paragraph{Graph terminology} Let~$G=(V,A)$ be a (un)directed graph.
We also use the notation~$V(G)$ and~$A(G)$ (resp.~$E(G)$) to denote the vertex set and arcs set (resp. edges set) of~$G$ respectively  
%The \emph{neighborhood} of a vertex~$v \in V$, denoted by~$N_G(v)$, is the set of all neighbors of~$v$. 
%The \emph{closed neighborhood} of a vertex~$v$, denoted~$N_G[v]$, is the set~$N_G(v) \cup \{v\}$. 
%The \emph{degree} of a vertex~$v$ is denoted by~$d_G(v)$. We may simply write~$N(v)$ and~$d(v)$ if the graph is clear from the context.
Let~$X \subseteq V$, we denote by~$G[X]$ the subgraph of~$G$ induced by~$X$.
%and by~$G - X$ the graph obtained from~$G$ by deleting the vertices in~$X$ and incident arcs.
We denote by~$N(v)$ the set of adjacent vertices of~$v$ and~$\delta^+(v)$ (resp.~$\delta^-(v)$) the set of outgoing (resp.~ingoing) arcs of~$v$.
A \emph{path} or $v_1v_\ell$-\emph{path} is a sequence of vertices $v_1 - v_2 - \ldots - v_\ell$ such that $v_iv_{i+1} \in A$ for each~${i=1,\ldots,\ell-1}$. If no vertices appear more than once in a path then it is \emph{elementary}.
The two vertices~$v_1$ and~$v_\ell$ are the \emph{endpoints} of the path and the others are called \emph{internal} vertices. A subpath is defined as any subsequence $v_j - v_{j+1} - \ldots - v_{j+k}$ for some~$j$ and~$k$. Unless stated otherwise, all the paths considered in this paper are elementary.
%We use the notation~$p_{uv}$ to denote a path with endpoints~$u$ and~$v$ and $p_{uv}[v_i, v_j]$ to denote a subpath of $p_{uv}$ with $v_i$, $v_j$ being two internal nodes.
We use the notation~$p[v_i, v_j]$ to denote a subpath of $p$ with $v_i$, $v_j$ being any two vertices of~$p$.
We denote by~$\mathcal{P}(G)$ the set of all elementary path in~$G$.

We say that~$G$ is \emph{bidirected} if $uv \in A$ and $vu \in A$ for all $u, v \in V$.
The \emph{underlying undirected graph}~$G^u$ of~$G$ is the undirected graph obtained from~$G$ by taking the 
same set of vertices, and with the set of edges defined as follows. There is an edge between any pair of vertices $u$ and $v$, if the directed graph has an arc $uv$ or $vu$.

%Given two paths~$p = u_1 - \ldots - u_\ell$ and~$q = v_1 - \ldots - v_\ell$, we denote by $p \oplus q$ the new path~$u_1 - \ldots - u_\ell - v_1 - \ldots - v_\ell$. 
%If no edges (resp. vertices) appear more than once in a path then it is \textsl{simple} (resp. \textsl{elementary}).

\paragraph{Tree decomposition and treewidth} A tree decomposition~$\mathcal{T} = (T, \mathcal{B})$ of an undirected graph~$G = (V, E)$ consists of a tree
$T = (X, F)$ with node set~$X$ and edge set~$F$, and a set~$\mathcal{B} \subseteq 2^V$ whose members~$B_x \in \mathcal{B}$, called \textsl{bags}, are labeled with the node~$x \in X$, such that the following conditions are met:

\begin{enumerate}
\item $\bigcup_{x \in X} B_x = V$.
\item For each~$uv \in E$ there is an~$x \in X$ with~$u, v \in B_x$.
\item For each~$v \in V$, the node set~$\{x \in X : v \in B_x \}$ induces a subtree of~$T$.
\end{enumerate}

The third condition is equivalent to assuming that if~$v \in B_{x'}$ and~$v \in B_{x''}$ then~$v \in B_x$ holds for every node~$x$ of the (unique)~$x'x''$-path in~$T$. The width of a tree decomposition~$\mathcal{T}$
is~$w(\mathcal{T}) = \max_{x \in X} |B_x | - 1$ and the treewidth of~$G$ is defined as~$\tw(G) = \min_{\mathcal{T}} w(\mathcal{T})$
where the minimum is taken over all tree decompositions~$\mathcal{T} = (T, \mathcal{B})$ of~$G$. The ``$-1$'' in the
definition of~$w(\mathcal{T})$ is included for the convenience that trees have treewidth~$1$ (rather than~$2$).

Any tree decomposition~$\mathcal{T} = (T, \mathcal{B})$ of a graph can be transformed in linear time into a
so-called nice tree decomposition $\mathcal{T}'= (T' , \mathcal{B}')$ with~$w(\mathcal{T}') = w(\mathcal{T})$,~$|\mathcal{B}'| = O(|\mathcal{B}|)$ and with~$B_x \neq \emptyset$ for all~$B_x \in \mathcal{B}$ where~$T'$ is a rooted tree satisfying the following conditions (see~\cite{kloks94} for more details):

\begin{enumerate}
\item Each node of~$T'$ has at most two children.
\item For each node~$x$ with two children~$y, z$, we have~$B_y' = B_{z}' = B_x'$ ($x$ is called \textit{join node}) with~$B_x', B_{y}', B_z' \in \mathcal{B}'$.
\item If a node~$x$ has just one child~$y$, then~$B_x' \subset B_y'$ ($x$ is called \textit{forget node}) or~$B_y' \subset B_x'$ ($x$ is called \textit{insert node}) and~$| |B_x' | - |B_y' | | = 1$ with $B_x', B_y' \in \mathcal{B}'$.
\end{enumerate}

One can see that the subtree~$T_x$ of~$T$ rooted
at node~$x$ represents the subgraph~$G_x$ induced by precisely those vertices of~$G$ which occur in at least one~$B_y$ where~$y$ runs over the nodes of~$T_x$. When the graph is directed, the tree decomposition applies for the underlying undirected graph.

\paragraph{Parameterized complexity} 
The parameterized complexity theory is a framework that provides a new way to express the computational complexity of optimization problems. We briefly recall here the main ideas behind this theory, the reader is referred to~\cite{DowneyF13} for more background on this subject. A problem parameterized by~$k$ is called \emph{fixed-parameter tractable} (fpt) if there exists an algorithm, called an fpt algorithm, that solves it in time~$f(k) \cdot n^{O(1)}$ (fpt-time) where~$n$ is the size of the input. The function~$f$ is typically super-polynomial and only depends on~$k$. In other words, the combinatorial explosion is confined into~$f$. If the values of~$k$ are small in practice, then the algorithm adopts a polynomial behavior. 

%A tree decomposition comes with the following important property

%\begin{lemma}
%\label{lem:td-sep}
%Let~~$\mathcal{T} = (T, \mathcal{B})$ be a tree decomposition. Assume that the deletion of a node~$x$ from~$T$ creates~$\ell$ subtrees~$T_1, \ldots, T_\ell$. Then the subgraphs~$G_1 - B_1,\ldots,G_\ell - B_\ell$ share no vertices and there are no edges connecting them.
%\end{lemma}

%This ``separation'' property is essential to solve the \minconuspr problem by dynamic programming as we will see in the next section.

%% file: problem_description.tex
\subsection{Notations and definitions}
In terms of graphs, the problem can be presented as follows. We consider given a bidirected graph $G$ = ($V$, $A$) that represents an IP network topology. Every node $v\in V$ corresponds to a router while an arc $a$ = $uv\in A$ represents a logical link between router nodes $u$ and $v$. Every arc $uv$ is associated a capacity (bandwidth) denoted by $c_{uv} \geqslant$ 0 and a latency value denoted $\delta_{uv}\geqslant$ 0. We let $K$ denote a set of commodities (traffic demands) to be routed over the graph $G$. Every commodity $k$ is defined by a pair ($s^{k}$, $t^{k}$) with $s^{k}$, $t^{k}$ being the origin and destination of $k$, respectively, along with the traffic volume $D^{k} \geqslant$ 0 to be routed from $s^{t}$ to $t^{k}$ and a a maximum delay value $\Delta^{k} \geqslant$ 0. 

\begin{definition}[Partial routing (sub)path]
A \textsl{partial routing path} for a commodity~$k \in K$ in a graph~$G$ is a pair~$(p, k)$ denoted by~$p^k$ where~$p$ is a path. A \textsl{partial routing subpath} of~$p^k$ is a routing path~$q^k$ such that~$q$ is a subpath of~$p$.
\end{definition}

\begin{definition}[Complete routing path]
A routing path~$p^k$ is said to be complete for a commodity~$k \in K$ in a graph~$G$ if the endpoints of~$p$ are exactly the origin and destination of~$k$.
\end{definition}

\begin{definition}[Routing configuration]
A \textsl{routing configuration} for a set of commodities~$K$ in a graph~$G$ is a subset~$R_{G,K} \subseteq \{p^k : p \in \mathcal{P}(G), k \in K\}$ of routing paths.
\end{definition}

The set of all possible routing configurations of~$K$ in~$G$ is denoted by~$\mathcal{R} (G, K)$.

\begin{definition}[Feasible routing configuration]
A routing configuration~$R \in \mathcal{R} (G, K)$ is said to be \textsl{feasible} if there exists a \textsl{weight function} $w:A \to \mathbb{Z}_{+}$ such that $(i)$ for each~${p^k \in R}$ the path~$p$ is the unique shortest path between its endpoints according to~$w$ and $(ii)$~the delay constraint is satisfied \emph{i.e}~$\sum_{p^k \in R} \sum_{uv \in p} \delta_{uv} \leqslant \Delta^{k}$.
\end{definition}

\begin{definition}[Complete routing configuration]
A routing configuration~$R \in \mathcal{R} (G, K)$ is said to be \textsl{complete} if it is feasible and there exists a (necessarily unique) complete routing path in~$R$ for each~$k \in K$.
\end{definition}

\begin{definition}[Conflicting paths]
Two paths $p_1$ and $p_2$ are said to be \textit{conflicting} if they share two vertices $u$ and $v$ and $p_1[u,v] \neq p_{2}[u,v]$ with $p_1[u,v] \neq\emptyset$ and $p_2[u,v] \neq\emptyset$. Otherwise, they are said to satisfy Bellman property.
\end{definition}

Finally, we provide the following two metrics:

\begin{definition}[Load]
The \textsl{load} $\mbox{\texttt{load}}(R, u, v)$ of an arc~${uv \in A}$ given a routing configuration~${R \in \mathcal{R}(G, K)}$ is defined as

$$\mbox{\texttt{load}}(R, u, v) = \displaystyle{\frac{1}{c_{uv}} \sum_{p^k \in R : uv \in p} D^k}$$
The load is the ratio between the total flow that goes through an arc and the arc's capacity.
\end{definition}

\begin{definition}[Congestion]
The \textsl{congestion} $\mbox{\texttt{cong}}(R)$ of a routing configuration~$R \in \mathcal{R}(G, K)$ is defined as

$$\mbox{\texttt{cong}}(R) = \displaystyle{\max_{uv \in A} \mbox{\texttt{load}}(R, u, v)}$$
The congestion is then the maximum load over all arcs.
\end{definition}

\subsection{Properties}

In this section, we will state two lemmas that will be useful in the rest of the paper.

\begin{lemma}
\label{lem:no-conflict}
Let~$R \in \mathcal{R} (G, K)$ if~$R$ is feasible then it contains no conflicting routing paths.
\end{lemma}
\begin{IEEEproof}
Suppose, on the contrary, that~$R$ is feasible and yet contains two routing paths~$p_1^{k}, p_2^{k'}$ such that they share two vertices $u$ and $v$ and $p_1[u,v] \neq p_{2}[u,v]$ with $p_1[u,v] \neq\emptyset$ and $p_2[u,v] \neq\emptyset$. Since~$R$ is feasible then there exists a weight function such that~$p_1$ (resp.~$p_2$) is the unique shortest path between its endpoints with respect to~$w$.
By the Bellman principle, the subpath~$p_1[u,v]$ is a shortest between~$u$ and~$v$.
For the same reason, the subpath~$p_2[u,v]$ is a shortest between~$u$ and~$v$ which is different from~$p_1[u,v]$ by assumption. Hence, there are two different shortest paths to join the endpoints, denoted~$s$ and~$t$, of~$p_1$ i.e the path~$p_1$ itself and the one formed by the subpaths~$p_1[s,u]$,~$p_2[u,v]$ and~$p_1[v,t]$. This contradicts the unicity of~$p_1$.
\end{IEEEproof}

The next lemma shows that feasibility can checked in polynomial time.

\begin{lemma}[Benameur and Gourdin \cite{benameur03}]
\label{lem:pl}
Determining whether a routing configuration is feasible and returning the corresponding weight function, if any, can be done in polynomial time.
\end{lemma}

\subsection{Problem statement}

We are now in position to state the problem studied in this paper. The \textit{Delay Constrained Minimum Congestion} (D-USPR) problem is to find a set of weights to assign to the arcs of $G$ and a set of routing paths induced by those weights such that $(i)$ there is a unique shortest path satisfying the delay constraints for each commodity according to the identified weights and $(ii)$ the network congestion is minimum.
%\vspace*{-0.53cm}
Formely, the problem is defined as follows.

\medskip

\problemopt{\duspr}
{A bidirected graph~$G=(V,A, c, \delta)$ where each arc~$uv \in A$ has a \emph{capacity} value~$c_{uv} \geqslant 0$ and a \emph{latency} value~$\delta_{uv}\geqslant 0$, a set~$K$ of commodities where each commodity~$k \in K$ is defined as~$(s^{k}$, $t^{k}$, $D^{k}$, $\Delta^{k})$.}
{A complete routing configuration~$R_{G,K}$ with minimum congestion value.}

\medskip

In this paper, we will also make use of this slightly more general version of the above problem

\medskip

\problemopt{\kduspr}
{A bidirected graph~$G=(V,A, c, \delta)$ where each arc~$uv \in A$ has a \emph{capacity} value~$c_{uv} \geqslant 0$ and a \emph{latency} value~$\delta_{uv}\geqslant 0$, a set~$K$ of commodities partitioned into two sets~$K_{free}$ (free demands) and~$K_{fixed}$ (fixed demands) and a complete routing configuration~$R_{G,K_{fixed}}$.}
{A complete routing configuration~$R_{G, K_{free}}$ such that~$R_{K_{free}} \cup R_{K_{fixed}}$ is feasible and has minimum congestion value.}

\medskip

Observe that if~$K_{fixed} = \emptyset$, we end up with the \duspr problem.

\begin{figure}
\centering
  \begin{tikzpicture}[transform shape, auto,swap]
  \begin{scope}
    \foreach \pos/ \thr / \name in {{(0,0)/b}, {(0,1)/a}, {(1,0.5)/c},
                            {(2,0)/e}, {(2,1)/d}, {(3,0.5)/f}}
        \node[vertex,label=\name] (\name) at \pos {};

    \foreach \source/ \dest in {b/c,a/c,c/e,c/d,d/f,e/f}
        \path[edge] (\source) -- (\dest);
\end{scope}
\begin{scope}[xshift=5cm]
    \foreach \pos/ \thr / \name in {{(0,0)/b}, {(0,1)/a}, {(1,0.5)/c},
                            {(2,0)/e}, {(2,1)/d}}
        \node[vertex,label=\name] (\name) at \pos {};

    \foreach \source/ \dest in {b/c,a/c,c/e,c/d}
        \path[edge,label=$1$] (\source) -- (\dest);
\end{scope}
\end{tikzpicture}
  %\caption{}
  %\label{fig:sub2}
\caption{In the graph on the left, the two paths~${a-c-d-f}$ and~${b-c-e-f}$ form the unique valid solution for the \edp problem where the demands are~$(b,f)$ and~$(a,f)$. However, this is not a feasible routing configuration since these two paths are conflicting. In the graph on the right, it is not possible to find two node-disjoint paths satisfying the demands~$(b,e)$ and~$(a,d)$, yet it is easy to see that the routing paths~$a-c-d$ and~$b-c-e$ form a feasible routing configuration.}
\label{fig:not-special-case}
\end{figure}
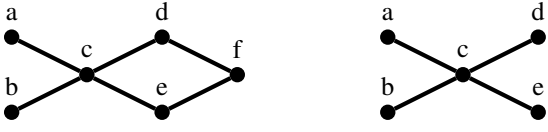

%%%%%%%% SPACE CONSTRAINT
It is worth noting that, even if they look similar at first glance, the \edp (edp) problem (resp. \ndp (ndp) problem) is not a particular case of \duspr with unit demands and unit capacities (see~Fig.~\ref{fig:not-special-case}). Recall that the edp (resp. ndp) problem asks, given an undirected graph and a set of~$k$ demands, to find~$k$ edge-disjoint (resp. node-disjoint) paths joining the demands. Consequently, the negative and positive results for edp or ndp do not directly transfer to \duspr.

%%%%%%%% SPACE CONSTRAINT

%% file: milp_formulation.tex
\subsection{Notations and formulation}
Let $x^{k}_{a}$ be a binary variable that takes the value 1 if commodity $k$ is routed along a path using arc $a$ and 0 otherwise. We define the binary variables $u^{t}_{a}$ that takes the value 1 if $a$ belongs to a shortest path towards destination $t$ and 0 otherwise. We further let $w_{uv}$ denote the weight assigned to the arc $uv$ and $r^{v}_{u}$ be the potential of node $u$, that is the distance between node $u$ and node $v$. The D-USPR problem is then equivalent to the following MILP formulation:  
\begin{alignat}{2}
\scriptsize \label{ctn:obj} & \min L\\
\label{ctn:flow} s.t. & \sum_{a \in \delta^+(v)} x^{k}_{a} - \sum_{a \in \delta^-(v)} x^{k}_{a} =  
\left \{ \begin{array}{ll}
1 & \mbox{if $v=s^{k}$,}\\
-1 & \mbox{if $v=t^{k}$,}\\
0 & \mbox{otherwise.}
\end{array} \right.
\begin{array}{l}\forall v \in V,\\ \forall k \in K, \end{array}\\
\label{ctn:load} & \sum_{k \in K} D^{k} x^{k}_{a}\leqslant c_{uv} L,   \forall a \in A,\\
\label{ctn:delay} & \sum_{a \in A} \delta_{a} x^{k}_{a}\leqslant \Delta^{k},  \forall k \in K,\\
\label{ctn:antiarborescence} & \sum_{a \in \delta^+(v)} u^{t}_{a}\leqslant 1,  \forall v \in V, \forall  t \in T,\\
\label{ctn:linking_1} & x^{k}_{a} \leqslant u^{t^{k}}_{a}, \forall a \in A, \forall k \in K, \\
\label{ctn:linking_2} & u^{t}_{a} \leqslant \sum_{k\in K, t^{k}=t}x^{k}_{a}, \forall a \in A, \forall t\in T,\\
\label{ctn:compatibility_1} & w_{uv} - r^{t}_u + r^{t}_v \geqslant 1 - u^{t}_{uv}, \forall uv \in A, \forall t\in T, \\
\label{ctn:compatibility_2} & w_{uv} - r^{t}_u +  r^{t}_v \leqslant M(1 - u^{t}_{uv}), \forall uv \in A, \forall t\in T, \\
%\label{ctn:trivial_potential_1} & r^{v}_v = 0,  & \forall v \in V,\\
%\label{ctn:trivial_potential_2} & r^{t^{k}}_{s^{k}}\geqslant 1, & \forall k \in K,\\
\label{ctn:trivial_x} & x^{k}_{a}\in\{0, 1\}, \forall k\in K, \forall a \in A,\\
\label{ctn:trivial_u} & u^{t}_{a}\in\{0, 1\}, \forall a \in A, \forall t\in T, \\
\label{ctn:trivial_w} & w_{uv} \geqslant 0, \forall uv \in A,\\
\label{ctn:trivial_r} & r^{v}_{u}\geqslant 0, \forall u, v \in V\times V.
\end{alignat}

The objective \eqref{ctn:obj} is to minimize the load of the most loaded link, denoted $L$. Inequalities \eqref{ctn:flow} ensure that a unique path is associated to each commodity $k$ and \eqref{ctn:load} express the load over an arc $a$. Inequalities \eqref{ctn:delay} are the delay constraints over the routing paths while \eqref{ctn:antiarborescence} and \eqref{ctn:linking_1}-\eqref{ctn:linking_2} are anti-arborescence and linking constraints, respectively. In particular, inequalities \eqref{ctn:antiarborescence} ensure that there is at most one path traversing any node $v$ towards a given destination~${t\in T}$, which is necessarily implied by Bellman property. Constraints \eqref{ctn:compatibility_1} and \eqref{ctn:compatibility_2} guarantee that the weight of any arc used by a shortest path towards a destination $t$ corresponds to the difference of potentials between the end nodes of this arc and larger otherwise. Finally, \eqref{ctn:trivial_x}-\eqref{ctn:trivial_r} are the trivial and integrity constraints.

\begin{proposition}
The formulation \eqref{ctn:flow}-\eqref{ctn:trivial_u} is valid for the D-USPR problem. 
\end{proposition}

\begin{IEEEproof}
It is easy to see that any solution of D-USPR satisfies inequalities \eqref{ctn:flow}-\eqref{ctn:trivial_u}. To show the converse, let $(x, u, w, r)$ denote a solution of \eqref{ctn:flow}-\eqref{ctn:trivial_u} and consider two sets, say $Q^{k}$ and $S^{t}$ defined as follows: $Q^{k}$ = $\{a\in A: x^{k}_{a} = 1\}$, for each $k\in K$, with $Q$ = $\cup^{k\in K}Q^{k}$ and $S^{t}$ = $\{a\in A: u^{t}_{a} = 1\}$, for every $t\in T$.  Since $x\in\{0, 1\}^{K\times A}$ satisfies the flow conservation constraints \eqref{ctn:flow}, $Q^{k}$ clearly contains a unique routing $s^{k}t^{k}$-path for commodity $k$, which by \eqref{ctn:delay} also satisfies the delay constraints. Inequalities \eqref{ctn:flow}, \eqref{ctn:antiarborescence}, \eqref{ctn:linking_1} and \eqref{ctn:linking_2} ensure that the elements of $S^{t}$ form an anti-arborescence rooted at destination $t$ and consequently, contains paths that satisfy Bellman property.
Moreover, since $w\in\mathbb{R}^{+}$ and $r\in\mathbb{R}^{+}$ satisfy inequalities \eqref{ctn:compatibility_1} and \eqref{ctn:compatibility_2}, every set $Q^{k}$, $k\in K$, consists of arcs which are tight with respect to the weights $w$, thus containing a shortest $s^{k}t^{k}$-path.

Now suppose that there are two conflicting paths in $Q$, say $p^{1}$ and $p^{2}$ and let $s^{1}t^{1}$ and $s^{2}t^{2}$ be their respective endpoints. Denote by $u$ and $v$ two internal nodes of $p^{1}$ and $p^{2}$ such that $p^{1}[u,v] \neq p^{2}[u,v]$. Further assume that 
\begin{equation*}
\sum_{v_i v_j\in p^{1}}w_{v_i v_j} = \sum_{u_i u_j\in p^{2}}w_{u_i u_j},
\end{equation*}
then the inequalities \eqref{ctn:compatibility_1} and \eqref{ctn:compatibility_2} with $t\in\{t_{1}, t_{2}\}$ summed over $v_i v_j\in p^{1}[u,v]$ and $u_i u_j\in p^{2}[u,v]$ yields a contradiction. Consequently, every $st$-path used for routing is a unique shortest path according to the weights $w$ and hence $(x,u,w,r)$ is a solution of the D-USPR problem. 
\end{IEEEproof}

\subsection{Valid inequalities}
In what follows, we present two families of inequalities valid for the D-USPR problem.\\

\textit{(i) Subpath consistency constraints} \\ 

The first family is the so-called \textit{subpath consistency constraints} and has been introduced in different versions in \cite{} 
\begin{proposition}[\cite{Bley2011}]
The following inequalities 
\begin{equation}
\label{eq:subpath_1} x^{s,v}_{a} - x^{s, t}_{a} + \sum_{e\in\delta^{-}(v)}x^{s, t}_{e} \leqslant 1, \quad \forall (s, t), (s, v) \in K, \forall a\in A,
\end{equation}
\begin{equation}
\label{eq:subpath_2} x^{v, t}_{a} - x^{s, t}_{a} + \sum_{e\in\delta^{-}(v)}x^{s, t}_{e} \leqslant 1, \quad \forall (s, t), (v, t) \in K, \forall a\in A,
\end{equation}
are valid for the D-USPR problem.
\end{proposition}
Those inequalities ensure that two paths $p^{1}$, $p^{2}$ with a common endpoint $v_{i}\in V$, that intersect at a second node $v_{j}$ are necessarily such that $p^{1}[v_{i}, v_{j}]$ = $p^{2}[v_{i}, v_{j}]$. In other words, any pair of commodities having the same origin (respectively destination) node are necessarily routed along two paths that satisfy Bellman property.\\

\textit{(ii) Node precedence constraints} \\

The second family of inequalities has been introduced by Garcia \cite{Garcia2009} for the \textit{resource constrained shortest path} problem and later extended by Horvath et al. \cite{Horvath2016}. They arise directly from the maximum delay requirement of the D-USPR problem and express the fact that any feasible routing $st$-path using an arc $a = uv\in A$ should leave this arc through an arc $a' = u'v'\in\delta^{+}(v)$ satisfying the following condition 
\begin{equation*}
    \sigma_{s, u} + \delta_{a} + \delta_{a'} + \sigma_{v',t} \leqslant \Delta^{s, t},
\end{equation*}
where $\sigma_{s, u}$ (respectively $\sigma_{v', t}$) is the length of the shortest path between nodes $s$ and $u$ (respectively between nodes $v'$ and $t$) with respect to the delay metric. In other words, $\sigma_{uv}$ = $\sum_{e\in p^{uv}}\delta_{e}$, for $(u, v)\in V\times V$. For each arc $a = uv\in A$ and each commodity $k\in K$ originating from node $s^{k}$ and going to node $t^{k}$, denote by $\Phi^{out}_{a, k}$, the set of arcs defined as follows 

\begin{equation*}
    \begin{split}
    \Phi^{out}_{a, k} = \{a' = u'v'\in\delta^{+}(v) \text{ with } v'\neq u: \\
    \sigma_{s^{t}, u} + \delta_{a} + \delta_{a'} + \sigma_{v',t^{k}} \leqslant \Delta^{k}\}.
    \end{split}
\end{equation*}
Similarly, we let $\Phi^{in}_{a, k}$ denote the set of arcs entering into node $u$ that can belong to a feasible routing path:
\begin{equation*}
\begin{split}
     \Phi^{in}_{a, k} = \{a' = u'v'\in\delta^{+}(u) \text{ with } u'\neq v: \\  \sigma_{s^{t}, u'} + \delta_{a'} + \delta_{a} + \sigma_{v, t^{k}} \leqslant \Delta^{k}\}.
     \end{split}
\end{equation*}
\begin{proposition}
The following inequalities
\begin{equation}
  \label{eq:node_prec_1}  x^{k}_{a} \leqslant \sum_{a'\in\Phi^{out}_{a, k}}x^{k}_{a, k}, \forall a\in A, \forall k\in K,
\end{equation}
\begin{equation}
  \label{eq:node_prec_2} x^{k}_{a} \leqslant \sum_{a'\in\Phi^{in}_{a, k}}x^{k}_{a, k}, \forall a\in A, \forall k\in K,
\end{equation}
are valid for the D-USPR problem.
\end{proposition}

\begin{IEEEproof}
Let $k$ be a commodity of $K$ with origin $s^{k}$ and destination $t^{k}$ and let $a = uv$ be an arc of $A$ such that $v \neq s^{k}, t^{k}$. Denote by ($\overline{x}$, $\overline{u}$, $\overline{w}$, $\overline{r}$, $\overline{L}$) and let $p^{k}$ = $\{e\in A: \overline{x}^{k}_{e} = 1\}$ be the $s^{k}t^{k}$-path associated with the routing of commodity $k$. It is easy to see that inequality \eqref{eq:node_prec_1} is trivially satisfied if $\overline{x}^{k}_{a}$ = 0. Now if $\overline{x}^{k}_{a}$ = 1, then, by \eqref{ctn:flow}, there exists an arc, say $a'$ = $u'v'$ in $\delta^{+}(v)$ such that $\sum_{e\in\delta^{+}(v)}\overline{x}^{k}_{a}$ = 1. Denote by $p[s^{k}u]$ = $\{e\in A: \overline{x}^{s^{k}u}_{e} = 1\}$ (respectively $p[v' t^{k}]$= $\{e\in A: \overline{x}^{v' t^{k}}_{e} = 1\}$) the subpath with endpoints $s^{k}, u$ (respectively, $v', t^{k}$. Suppose that $a'$ is in $\delta^{+}(v)\setminus\Phi^{out}_{a, k}$, that is to say,  
\begin{equation*}
    \sum_{e\in p[s^{k},u]}\delta_{e} + \delta_{a} + \delta_{a'} + \sum_{e\in p[v', t^{k}]}\delta_{e} \geqslant
\end{equation*}
\begin{equation*}
    \sigma_{s^{k}, u} + \delta_{a} + \delta_{a'} + \sigma_{v',t^{k}} > \Delta^{k},
\end{equation*}
which, by \eqref{ctn:delay}, yields a contradiction. Thus \eqref{eq:node_prec_1} are valid for D-USPR problem. 
We use similar arguments to show that inequalities \eqref{eq:node_prec_2} are valid for the problem. 
\end{IEEEproof}

%% file: iterative_algorithm.tex
In this section, we introduce an effective iterative algorithm for solving the D-USPR problem. Consider given a graph~$G$ with a capacity vector $C$ and a set of demands $K$ with a latency vector $\Delta$. The idea of this algorithm is to iteratively decrease the load by constructing feasible routing configurations and the associated weight vectors. To this end, we perform the following initialization steps:

\textbf{Step 1} We first solve the \textit{minimum congestion spanning tree} with delay constraints in $G$. Let~$H$ denote the spanning tree obtained. 

\textbf{Step 2} We set an arbitrary weight value $w^{0}_{uv}$ on each arc~$uv$ of $A(H)$ and a infinite weight on the remaining arcs~${uv \in A(G)\setminus A(H)}$. 

\textbf{Step 3} We associate with each commodity of $K$ a path, say~$p^{k}$, in~$H$ between~$s^{k}$ and~$t^{k}$. 

We will denote by $R_{0}$ the complete routing configuration obtained at the end of the initialization phase. Note that $R_{0}$ obviously defines a feasible solution for the D-USPR problem.  

We let~$a^{*}\in A(G)$ be the most loaded arc with respect to~$R_{0}$, that is to say $a^{*}$ = $\arg\max_{uv\in A} \mbox{\texttt{load}}(R, u, v)$. At each iteration, we then apply the following procedure. We consider a partition of the demands set~$K$ into two subsets~$K_{fixed}$ and~$K_{free}$, where~$K_{free}$ is the subset of (\textit{congesting}) demands whose routing path in~$R_{0}$ uses the arc~$a^{*}$ and~$K_{fixed}$ contains the remaining demands. We fix the routing paths of the demands in~$K_{fixed}$ along with the associated weights. Let~$T_{fixed}$ (resp.~$T_{free}$) denote the destination nodes of the demands in~$K_{fixed}$ (resp.~$K_{free}$) and~$R^{t} \subseteq R_{K_{fixed}}$ denotes this fixed routing toward destination~${t\in T_{fixed}}$. We determine a feasible routing configuration by rerouting the demands of~$K_{free}$, all other demands remaining equal. This can be done by solving the formulation \eqref{ctn:obj}-\eqref{ctn:trivial_r} with the following changes. Inequalities \eqref{ctn:flow}, \eqref{ctn:delay}-\eqref{ctn:compatibility_2} are written over~$K_{free}$ and~$T_{free}$ instead of~$K$ and~$T$ while \eqref{ctn:load} is replaced by the following inequality
\begin{equation}
\label{ctn:load_bis} \sum_{k \in K_{free}} D^{k} x^{k}_{a} + \sum_{k \in K_{fixed}} D^{k} \overline{x}^{k}_{a}\leqslant c_{uv} L,  \forall a \in A,    
\end{equation}
Finally, we add the following inequalities 
\begin{align}
\label{ctn:compatibility_1_bis} & w_{uv} - r^{t}_u + r^{t}_v = 0, \quad \forall uv \in R^{t}, \forall t\in T_{fixed}, \\
\label{ctn:compatibility_2_bis} & w_{uv} - r^{t}_u +  r^{t}_v \geqslant 1, \quad \forall uv \in A\setminus R^{t}, \forall t\in T_{fixed},
\end{align} 
to ensure that the paths fixed in each set $R^{t}$ define shortest paths towards the destination $t\in T_{fixed}$. This procedure is summarized in Algorithm \ref{alg:iterative_algorithm}.

\begin{algorithm}[]
	\KwData{An instance ($G$, $K$, $C$, $\Delta$) of the problem}
	\KwResult{A complete routing configuration $R$}
 \Init{
 $R^0 \leftarrow$ a complete routing configuration obtained by performing Step1-Step3 \\
 $a^{*}\leftarrow \arg\max_{a\in A}\mbox{\texttt{cong}}(R^{0}, a)$ \\
 $iter \leftarrow$ 0 
}

	\While{$iter < iter_{max}$}{
		    find the demands in $K_{free}$ and $K_{fixed}$\\ 
		    find a complete routing configuration ${R^{iter}_{G, K_{fixed}}}$ \\
		     $R_{iter} \leftarrow$ complete routing configuration obtained by solving \eqref{ctn:obj}- \eqref{ctn:compatibility_2_bis} \\ 
		    $a^{*} \leftarrow \arg\max_{a\in A} \mbox{\texttt{load}}(R_{iter}, a)$  \\
		    iter $\leftarrow $ iter + 1
		    
	}
		\Return $R_{iter}$ with ${\mbox{\texttt{cong}}(R) \leqslant \mbox{\texttt{cong}}(R_0)}$\;

\caption{Iterative algorithm}
\label{alg:iterative_algorithm}
\end{algorithm}

%% file: dynamic_prog.tex
In this section, we introduce a dynamic programming algorithm based on a \textit{tree decomposition} for solving the \kduspr problem.
%In this section, we will present our dynamic programming algorithm for solving the~\minconuspr~problem in fpt-time with respect to the combined parameter ``treewidth'' and the ``number of demands''.
Observe that the problem is trivial in the case where the input graph is a tree since there can only be one path to route any demand. However, it is not possible to generalize this positive result to graphs of bounded treewidth since the problem is NP-complete even on bidirected rings~\cite{bley2009}. This negative result rules out the possibility of having an fpt algorithm parameterized only by the ``treewidth''. However, we prove in what follows that the problem is fixed-parameter tractable for the combined parameter ``treewidth'' and ``number of demands''.
%This negative result rules out the possibility of having an fpt algorithm parameterized only by the ``treewidth''. 
%Furthermore, since the \edp problem is a special case of \minunitconuspr and has been proven W[1]-hard w.r.t the ``treewidth'' \cite{ganian17}, then the same negative result also holds for the \minunitconuspr problem.
%As in the previous section, let $(G=(V,A,c,\delta)$, $K)$ be an instance of \duspr. We partition the set~$K$ into two sets~$K_{free}$ and~$K_{fixed}$.

\begin{proposition} \label{th:twunit}
Given a nice tree decomposition of~$G^u$  of width~$\omega$, the \kduspr problem can be solved in at most \dynalgotimek-time where~${\Delta = \max_{k \in K} \Delta^k}$.
\end{proposition}

\begin{IEEEproof}
Let ${I = (G=(V,A,c,\delta), K_{free}, K_{fixed}, R_{K_{fixed}})}$ be an instance of \kduspr.  
Let~$\mathcal{T} = (T = (X, F), \mathcal{B})$ be a nice tree decomposition of~$G^u$ rooted at node~$r \in X$.
We denote by~$\omega$ the width of~$\mathcal{T}$ and by~$n$ the order of~$G$ i.e~${n = |V|}$. 
We start the proof by introducing some extra notations and definitions.
%To make the exposition simpler and since we are dealing with bidirected graphs or subgraphs, we will reason on the graph~$G^u = (V, E)$ instead of~$G$ directly i.e. when a demand is routed between two vertices we do not precise the direction. In order to alleviate the notation, we drop the~$^u$ in~$G^u$.

\textbf{Notations.} Recall that~$T_x$ is the subtree of~$T$ rooted
at node~$x$ and~${G_x = (V_x, A_x)}$ is the subgraph of~$G$ induced by the vertices of~$G$ which occur in at least one bag~$B_y$ where~$y$ runs over the nodes of~$T_x$. In this proof, we will also use the subgraph~$\bar{G}_x$ which is obtained from~$G_x$ by removing the arcs with both endpoints in~$B_x$. 
We denote by~$U_x$ the set of all origins and destinations of the demands in~$K_{free}$ that lie in~$V_x$ i.e $U_x = \{ \{s^k, t^k\} \cap V_x: k \in K_{free} \}$  (see Figure~\ref{fig:algo-illus}). Let~$R \in \mathcal{R} (G, K)$ and~$G'$ a subgraph of~$G$, we denote by~$R\vert_{G'}$ the routing configuration obtained by taking the subpaths of~$R$ induced by~$G'$. We denote by~$G_R$ the graph obtained by taking the union of all routing paths in~$R$.

\textbf{Definitions.} In this paragraph, we introduce several notions that are needed in the proof. 

\textit{Valid routing configuration:}  We say that a routing configuration~${R \in \mathcal{R} (\bar{G}_x, K)}$ is \emph{valid}, if it is feasible and for every~$k \in K$ one of the following two conditions is met:
\begin{itemize}
    \item There is exacly one complete routing subpath~$p^k \in R$.
    \item The graph induced by the union of the routing subpaths for~$k$ in~$R$ is made of disjoint paths with at least one endpoint in~$B_x$. Furthermore, there are at most two degree-one vertices in~$V_x \setminus B_x$ in such graph.
\end{itemize}
%has at least one endpoint in~$B_x$. 

%We may assume in the remaining proof that all the routing configurations in~$\mathcal{R} (\bar{G}_x, K)$ are valid.
%Indeed, it is easy to see that any routing configuration in~$\bar{G}_x$ induced by some complete routing configuration in~$\mathcal{R} (G, K)$ must be valid.

\begin{figure}[!ht]
\centering
\begin{tikzpicture}[transform shape, auto,swap]
\begin{scope}[yshift=2cm]
		\node[terminal, red!80, label=left:a] (a) at (0,0) {};
		\node[terminal, red!80, label=below:b] (b) at (1,-2) {};
		\node[vertex, red!80, label=right:c] (c) at (2,0) {};
		\node[vertex, red!80, label=left:d] (d) at (0,2) {};
		\node[terminal, label=right:e] (e) at (2,2) {};
		\node[terminal, label=f] (f) at (0,4) {};
		\node[vertex, label=g] (g) at (2,4) {};

		\node (txt) at (0,-2.5) {$G_x$};
    	\node (txt) at (2,-2.5) {\textcolor{red!80}{$\bar{G}_x$}};
	
		\node (txt) at (1,-3.5) {$G$};

        \foreach \source/ \dest in {f/d, d/a, d/e, c/e, e/g}
            \path[edge, -] (\source) -- (\dest);        

        \foreach \source/ \dest in {b/c, a/b}
            \path[edge, -, red!80] (\source) -- (\dest);

        \begin{pgfonlayer}{background}
    		\path[fill=blue!20,rounded corners]
                (-0.5,-0.1) rectangle (2.6,-2.8);
			\draw[fill=blue!20,rounded corners, ultra thick, dashed] (-0.5,-0.4) -- (-0.5,2.6) -- (0.5,2.6)  -- (0.5,0.4) -- (2.6,0.4) -- (2.6,-0.4) -- cycle;
        \end{pgfonlayer}

\end{scope}
\begin{scope}[xshift=5cm, yshift=-0.5cm]
	\node[bag] (n1) at (0,0) {$a, b, c$};
	\node[bag] (n2) at (0,1) {$a, c$};
	\node[bag, ultra thick, dashed] (n3) at (0,2) {$a, c, d$};
	\node[bag] (n4) at (0,3) {$c, d$};
	\node[bag] (n5) at (0,4) {$c, d, e$};
	\node[bag] (n6) at (0,5) {$d, e$};
	\node[bag] (n7) at (2,5) {$d, e$}; %
	\node[bag] (n8) at (2,4) {$d$}; %
	\node[bag] (n9) at (1,6) {$d, e$};
	\node[bag] (n10) at (1,7) {$e$};
	\node[bag] (n11) at (2,3) {$d, f$}; %
	\node[bag] (n12) at (1,8) {$e, g$};

    \node (txt-1) at (0.2,8) {$r$};
    \node (txt-2) at (-1,2) {$x$};
    \node (txt-3) at (1,-1) {$T$};

    \path[edge] (n1) -- (n2) -- (n3) -- (n4) -- (n5) -- (n6);
    \path[edge] (n9) -- (n7) -- (n8) -- (n11);
    \path[edge] (n6) -- (n9) -- (n10) -- (n12);
\end{scope}
\end{tikzpicture}
\caption{Example of a graph~$G$ together with a nice tree decomposition~${\mathcal{T} = (T=(X,F), \mathcal{H})}$ of~$G^u$ rooted at node~${r \in X}$. In order to alleviate the figure and since the graph is bidirected, we drop the arcs orientation. We have~$B_x = \{a,c,d\}$ and~$U_x = \{a, b\}$. The origins and destinations of demands are represented with squares.
%The routing configuration~$R_{K_{fixed}}$ contains only the routing path~$p^{k_3} = b-c-e$ represented as a continuous grey arrow. 
%The dashed grey arrows represent a feasible routing configuration~$R_{K_{free}}$ for~$K_{free}$.
%More precisely, the free demands~$(b,f)$, and~$(a,e)$ are routed along the two routing paths $p^{k_1} = b-a-d-f$ and~$p^{k_2} = a-d-e$, respectively. Observe that the routing configuration~$R_{K} = R_{K_{free}} \cup R_{K_{fixed}}$ is feasible and has congestion value~$2$.
%Furthermore,~$R_{K}$ induces a partial routing configuration that consists of three routing subpaths~$p^{k_1}_1 = b-a-d$,~$p^{k_2}_1 = a-d$, and~$p^{k_3}_1 = b-c$ in~$G_x$.The routing subpaths for the free demands are stored in the table~$A_x$ as $A_x[\rc] = \{b-a, a-d\}$ where $\rc(b,a) = \{(b,f)\}$ and $\rc(a,d) = \{(b,f), (a,e)\}$. The congestion value induced by $\rc$ is equal to~$2$.
}
\label{fig:algo-illus} 
\end{figure}
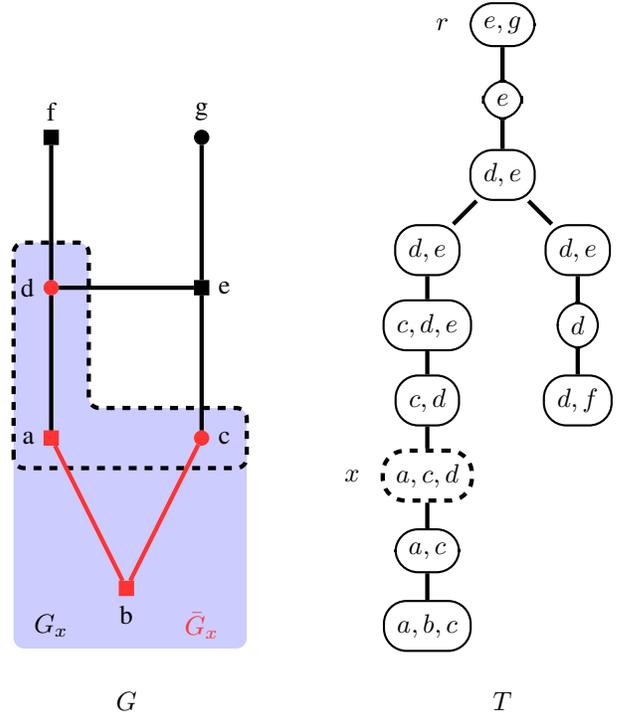

%We start by describing the subproblems solved by the dynamic programming algorithm. For this purpose, 

\textit{Routing contract:} A ``routing contract''~$H$ induced by a valid routing configuration~${R \in \mathcal{R} (\bar{G}_x, K)}$ is an edge-labeled graph where the edge labelling is a function~$\lambda_H$ from~$E(H)$ to~$2^{K}$ defined as follows.
%Let~$H'$ be the subgraph of~$\bar{G}_x$ obtained from the union of all the routing paths in~$\bar{R}_x$. 
First, we say that a vertex~${v \in V(G_{R})}$ is a \emph{transit} vertex if it has degree~$2$ in~$G_{R}$ and the demands routed along the edges~$v v_1$ and~$v v_2$ according to~$R$ where~${v_1, v_2 \in N(v)}$ are the same.
The graph~$H$ is then obtained from~$G_{R}$ by removing every transit vertex~${v \in V(G_{R})}$ and inserting the edge~$v_1 v_2$ i.e we remove~$v$ and connect its neighbors with an edge.
Regarding the edge labelling function~$\lambda_H$, the demands in~$\lambda_H(uv)$ are exactly those routed along the corresponding subpath~$p[u, v]$ (which can be a single edge) in $G_{R}$ (see Figure~\ref{fig:algo-subproblems-desc}). 
More generally, we say that a routing configuration~$Q  \in \mathcal{R} (\bar{G}_x, K)$ is $H$-respecting if there exists a mapping~$f : V(H) \to V(G_Q)$ such that for all~$uv \in E(H)$ the demands in~$\lambda_H(uv)$ are exactly those routed along the corresponding subpath~$p[f(u), f(v)]$ in~$G_Q$. We denote by~$\mathcal{H}_x$ the set of all possible routing contracts at node~$x$.

\begin{figure}[!ht]
\centering
\begin{tikzpicture}[transform shape, auto,swap]
\begin{scope}
	\node[vertex, label=above:a] (a) at (0,-1) {};
	\node[vertex, label=above:b] (b) at (0.75,-1) {};
	\node[vertex, label=above:c] (d) at (1.25,-1) {};
	\node[vertex, label=above:d] (e) at (2,-1) {};
	
	\node[vertex, label=left:e] (f) at (0.5,-2) {};
	\node[vertex, label=right:f] (g) at (1.5,-2) {};
	
	\node[vertex, label=below:g] (h) at (1,-3) {};
	
	\node (txt) at (1,-4) {$G$};
	
    \foreach \source/ \dest in {f/a, f/b, f/g, g/d,g/e,h/f,h/g, b/d}
        \path[edge, -] (\source) -- (\dest);
        
	\path[draw=blue!80, ultra thick, -] (1.05,-2.8) -- (0.2,-1);
	\path[draw=blue!80, ultra thick, -] (1.45,-1) -- (1.52,-1.75) -- (1.8,-1);
	
	\path[draw=green!50, ultra thick, -] (0.8,-3) -- (-0.2,-1);
	
	\path[draw=red!80, ultra thick, -] (1.7,-2) -- (2.2,-1);
\end{scope}
\begin{scope}[xshift=3cm]
	\node[vertex, label=above:a] (a) at (0,-1) {};
	\node[vertex, label=above:c] (d) at (1.25,-1) {};
	\node[vertex, label=above:d] (e) at (2,-1) {};
	
	\node[vertex, label=left:e] (f) at (0.5,-2) {};
	\node[vertex, label=right:f] (g) at (1.5,-2) {};
	
	\node[vertex, label=below:g] (h) at (1,-3) {};

	\node (txt) at (1,-4) {$G_{R}$};
	
    \foreach \source/ \dest in {f/a, g/d,g/e,h/f}
        \path[edge, -] (\source) -- (\dest);
        
	\path[draw=blue!80, ultra thick, -] (1.05,-2.8) -- (0.2,-1);
	\path[draw=blue!80, ultra thick, -] (1.45,-1) -- (1.52,-1.75) -- (1.8,-1);
	
	\path[draw=green!50, ultra thick, -] (0.8,-3) -- (-0.2,-1);
	
	\path[draw=red!80, ultra thick, -] (1.7,-2) -- (2.2,-1);
\end{scope}
\begin{scope}[xshift=6cm]
	\node[vertex, label=above:a] (a) at (0,-1) {};
	\node[vertex, label=above:c] (d) at (1.25,-1) {};
	\node[vertex, label=above:d] (e) at (2,-1) {};
	
	\node[vertex, label=right:f] (g) at (1.5,-2) {};
	
	\node[vertex, label=below:g] (h) at (1,-3) {};

	\node (txt) at (1,-4) {$H$};
	
    \foreach \source/ \dest in {h/a, g/d,g/e}
        \path[edge, -] (\source) -- (\dest);
        
	\path[draw=blue!80, ultra thick, -] (1.05,-2.8) -- (0.2,-1);
	\path[draw=blue!80, ultra thick, -] (1.45,-1) -- (1.52,-1.75) -- (1.8,-1);
	
	\path[draw=green!50, ultra thick, -] (0.8,-3) -- (-0.2,-1);
	
	\path[draw=red!80, ultra thick, -] (1.7,-2) -- (2.2,-1);
\end{scope}
\end{tikzpicture}
\caption{Illustration of the construction of a routing contract. In this example, there are three demands~$k_1$ (green), $k_2$ (blue) and~$k_3$ (red) routed according to a valid routing configuration~$R$. The different routing paths are represented as continuous colored line. The graph~$G_{R}$ is defined as the union of all routing paths in~$R$. The routing contract~$H$ of~$R$ is then obtained by replacing the only transit vertex~$e \in V(G_{R})$ by the edge~$ag$. Regarding the labeling function~$\lambda_{H}$ of~$H$, it is defined as follows:~${\lambda_{H}(ag) = \{k_1, k_2\}}$,~${\lambda_{H}(cf) = \{k_2\}}$ and~${\lambda_{H}(df) = \{k_2, k_3\}}$.
%The free demands are~${k_1 = (b,f,1,3)}$,~${k_2 = (a,e,1,2)}$ and the fixed demand is~${k_3 = (b,e,1,4)}$. The arcs have unit capacities and delays.
%The routing configuration~$R_{K_{fixed}}$ contains only the routing path~$p^{k_3} = b-c-e$ represented as a continuous grey arrow. 
%The dashed grey arrows represent a feasible routing configuration~$R_{K_{free}}$.
%More precisely, the free demands~$k_1$ and~$k_2$ are routed along the two routing paths $p^{k_1} = b-a-d-f$ and~$p^{k_2} = a-d-e$, respectively. Observe that the routing configuration~$R_{K} = R_{K_{free}} \cup R_{K_{fixed}}$ is feasible (e.g. set a weight of~$1$ to every arc) and has congestion value~$\mbox{\texttt{cong}}(R_K) = 2$.
%The associated routing contract of~$R_{K_{free}}$ is the function~$\rc$ defined as $\rc(\{a,b\}) = \{k_1\}$ and $\rc(\{a,c\}) = \rc(\{a,d\}) = \rc(\{b,c\}) = \rc(\{b,d\}) = \rc(\{c,d\}) = \emptyset$.
}
\label{fig:algo-subproblems-desc} 
\end{figure}
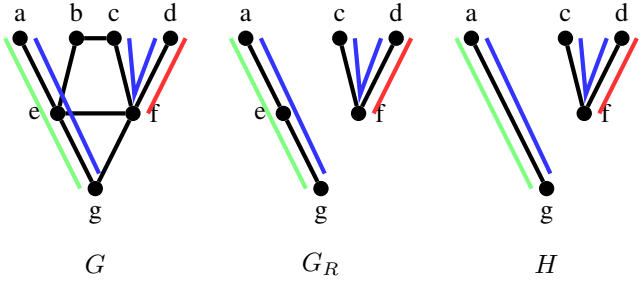

\textit{Delay contract:}  A ``delay contract'' induced by a valid routing configuration~${R \in \mathcal{R} (\bar{G}_x, K)}$ is a function~${d : K \to \mathbb{N}}$ defined as follows.
Given a demand~$k \in K$, the value~$d(k)$ is equal to the sum of the delays on the arcs used to route the demand~$k$ according to~$R$. We say that a routing configuration~$Q  \in \mathcal{R} (\bar{G}_x, K)$ is $d$-respecting if for each~${k \in K}$ we have~$\sum_{p^k \in Q} \sum_{uv \in p} \delta_{uv} = d(k)$. We denote by~$\mathcal{D}_x$ the set of all possible delay contracts at node~$x$.

\textbf{Subproblems definition.} We define a set of subproblems for each node~${x \in X}$, one corresponding to each possible~${H_x \in \mathcal{H}_x}$ and each~${d_x \in \mathcal{D}_x}$ that may represent in~$\bar{G}_x$ the routing contract and delay contract  induced by an optimal complete routing configuration for~$I$. Hence, for each routing contract~$\rc_x$ and each delay contract~$\dc_x$, we let~$OPT_x(\rc_x, \dc_x)$ be an $\rc_x$-respecting and $\dc_x$-respecting valid routing configuration in~$\mathcal{R}(\bar{G}_x, K)$ with minimum congestion. If no such routing configuration exists, we simply set~${OPT_x(\rc_x, \dc_x) = \emptyset}$.

\textbf{Recurrence relations.} 
%The algorithm starts by solving the subproblems associated to each leaf node~$\ell$ of $T$ (i.e the base cases). For each routing contract~$\rc_\ell$ and each routing intersection~$\dc_\ell$, we simply enumerate all feasible routing configurations in the subgraph~$G_\ell$ and stores in~$OPT_\ell(\rc_\ell, \dc_\ell)$ the one that is both~$\rc_\ell$-respecting and~$\dc_\ell$-respecting and has minimum congestion. The number of routing configurations in~$G_\ell$ is at most~$2^{|A_\ell||K_{free}|}$.
%Notice that there are no more than~$\omega + 1$ vertices in~$G_\ell$ thus the number of routing configurations is bounded by~$2^{(\omega + 1)\omega|K_{free}|}$.
We now describe how the solutions of the subproblems attached to a node are constructed. 
At the cost of adding more nodes in the tree~$T$, we may assume w.l.o.g that the bags associated to the leaves of~$T$ contains only one vertex. In this case, each leaf is considered as an insert node.
Initially,~$OPT_x(\rc_x, \dc_x) = \emptyset$ for all~$\rc_x, \dc_x \in \rcset_x \times \dcset_x$ and~$x \in X$.
By convention, $\mbox{\texttt{cong}}(\emptyset) = +\infty$.
%After solving the subproblems associated to leaves nodes as described above, 
The algorithm computes the table~$OPT_x$ of each node~$x$ in~$T$ according to their type (insert, forget or join) and using a bottom-up procedure that ends to the root as follows.

\paragraph{Insert node} Let~$x$ be an insert node. In the case that~$x$ is a leaf, we simply skip this step and move on to the next node. Otherwise, let~$y$ be the child of~$x$. 
By definition $B_y \subset B_x$ and~${B_x \setminus B_y = \{v\}}$.
We compute the table~$OPT_x$ as follows. For each~$\rc_y, \dc_y \in \rcset_y \times \dcset_y$ we perform the following instructions in sequence
\begin{itemize}
    \item We define a routing contract~$\rc_x$ obtained from~$\rc_y$ by simply adding the vertex~$v$.
    \item We construct a delay contract~$\dc_x$ by simply setting~${\dc_x= \dc_y}$.
\end{itemize}
After the instructions are performed, we set~${OPT_x(\rc_x, \dc_x) = OPT_y(\rc_y, \dc_y)}$.

\paragraph{Forget node} Let~$x$ be a forget node with child~$y$. By definition $B_x \subset B_y$ and~${B_y \setminus B_x = \{v\}}$.
Let~$E_v$ be the set of edges incident to~$v$ and~$d_{B_x}(v) = |N(v) \cap B_x|$.
This step requires the most attention since it is during this phase that we need to take care of the different ways to extend the routing paths of every routing configuration stored in~$OPT_y$ (See Figure~\ref{fig:algo-forget}). 
In what follows, we assume that whenever some fixed demands are routed along some of the edges in~$E_v$ we include the corresponding routing subpaths of~$R_{K_{fixed}}$ into every routing extension constructed hereafter.

\noindent
For each routing contract~$\rc_y \in \rcset_y$ and each delay contract~$\dc_y \in \dcset_y$ such that~${OPT_y(\rc_y, \dc_y) \neq \emptyset}$, we partition the set~$K_{free}$ into the following three sets:
\begin{itemize}
    \item $K_{open}$ : the free demands with no routing path in~$OPT_y(\rc_y, \dc_y)$.
    \item $K_{partial}$ : the free demands which have at least one (non-complete) routing path in~$OPT_y(\rc_y, \dc_y)$. 
    \item $K_{closed}$ : the free demands for which there exists a complete routing path in~$OPT_y(\rc_y, \dc_y)$.
\end{itemize}

First, we can ignore the demands in $K_{closed}$ : since they are end-to-end routed, there are no more decisions to be made for them. Consider instead a free demand~$k \in K_{open}$. Suppose for the moment that~$v \neq s^k$ and~$v \neq t^k$. Hence, this demand can be (possibly) routed through the vertex~$v$ using two edges of~$E_v$.
This yields to at most~$d_{B_x}(v) (d_{B_x}(v) - 1) / 2$ choices to route~$k$ through~$v$. Thus, a total of at most~$(d_{B_x}(v) (d_{B_x}(v) - 1) / 2)^{|K_{open}|}$ possibilities to route all the demands in~$K_{open}$ in this way. If~$v = s^k$ or~$v = t^k$ then the only choice is to pick one of the edge in~$E_v$ to start (or finish) routing the demand. Clearly, the number of possibilities in this case is dominated by the previous case.

Now consider a free demand~$k \in K_{partial}$. Let~$R_v^k$ be the set of routing paths for~$k$ in~$OPT_y(\rc_y, \dc_y)$ having at least one endpoint in~${(N(v) \cap B_x) \cup \{v\}}$. 
We will show how many new routing paths can be obtained to route~$k$ through~$v$ by extending those in~$R_v^k$.
%There are several cases to take into account. 
Similarly, suppose that~$v \neq s^k$ and~$v \neq t^k$. 
%If there are two paths~$p^k^1, p^k^2 \in R_v^k$ sharing~$v$ as endpoint, then the only choice is to concatenate at~$v$ these two paths into one. In the other cases, 
The demand~$k$ can be (possibly) routed through the vertex~$v$ by extending the paths in~$R_v^k$ with at most one or two edges of~$E_v$. Thus the total number of possible ways to extend the routing paths in~$R_v^k$ is bounded by~$(d_{B_x}(v) (d_{B_x}(v) - 1) / 2)$. Thus, a total of at most~$(d_{B_x}(v) (d_{B_x}(v) - 1) / 2)^{|K_{partial}|}$ possibilities to route all the demands in~$K_{partial}$.
Suppose now that~$v = s^k$ or~$v = t^k$, if there exists a routing path~$p^k \in R_v^k$ then the path cannot be extended through~$v$. Otherwise, the only choice is to extend the routing paths in~$R_v^k$ by picking one of the edge in~$E_v$ to start (or finish) routing the demand. Clearly, the number of possibilities in this case is dominated by the previous case.

Overall, there are at most
$${(d_{B_x}(v) (d_{B_x}(v) - 1) / 2)^{|K_{closed}| + |K_{partial}|}}$$ 
new possible routing configurations that can be constructed from the ones in~$OPT_y(\rc_y, \dc_y)$. Let~$\mathcal{R}_x$ be the set of those routing configurations that are valid (recall that checking whether a routing configuration is feasible can be done in polynomial time using Lemma~\ref{lem:pl}).
For each~${R_x \in \mathcal{R}_x}$, let~$\rc_x \in \rcset_x$ and~$\dc_x \in \dcset_x$ be the routing contract and delay contract induced by~$R_x$ (i.e $R_x$ is~$\rc_x$-respecting and~$\dc_x$-respecting), 
we set~$OPT_x(\rc_x,\dc_x) = R_x$ if~${\mbox{\texttt{cong}}(R_x) < \mbox{\texttt{cong}}(OPT_x(\rc_x, \dc_x))}$.
%we update the table~$OPT_x$ as follows :

%$$OPT_x(\rc_x) = \underset{R \in \{R_x, OPT_x(\rc_x)\} }{\argmin} \mbox{\texttt{cong}}(R) $$

%By definition of~$K_{partial}$, there must exist a (non-complete) routing path~$p^k \in OPT_y(\rc_y)$.

%and each demand~$k \in K_{free}$, perform the following steps in sequence
%\begin{itemize}
%    \item Let~$R_v \subseteq OPT_y(\rc_y)$ be the routing paths of~$OPT_y(\rc_y)$ having~$v$ as endpoint.
%    \item if~$v = s^k$ or~$v = t^k$
%\end{itemize}

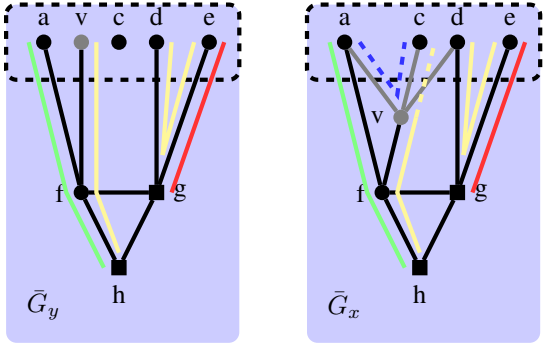
\begin{figure}[!ht]
\centering
\begin{tikzpicture}[transform shape, auto,swap]
\begin{scope}
		\node[vertex, label=above:a] (a) at (0,0) {};
		\node[vertex, black!50, label=above:v] (b) at (0.5,0) {};
		\node[vertex, label=above:c] (c) at (1,0) {};
		\node[vertex, label=above:d] (d) at (1.5,0) {};
		\node[vertex, label=above:e] (e) at (2.2,0) {};
		
		\node[vertex, label=left:f] (f) at (0.5,-2) {};
		\node[terminal, label=right:g] (g) at (1.5,-2) {};
		
		\node[terminal, label=below:h] (h) at (1,-3) {};
		
		\node (txt) at (0,-3.5) {$\bar{G}_y$};

        \foreach \source/ \dest in {f/a, f/b, f/g, g/d,g/e,h/f,h/g}
            \path[edge, -] (\source) -- (\dest);

		\path[draw=yellow!50, ultra thick, -] (1,-2.8) -- (0.7,-2) -- (0.7,0);
		\path[draw=yellow!50, ultra thick, -] (1.7,0) -- (1.58,-1.5) -- (2,0);
		
		\path[draw=green!50, ultra thick, -] (0.8,-3) -- (0.3,-2) -- (-0.2,0);
		
		\path[draw=red!80, ultra thick, -] (1.7,-2) -- (2.4,0);

        \begin{pgfonlayer}{background}
    		\path[fill=blue!20,rounded corners]
                (-0.5,-0.1) rectangle (2.6,-4);
			\draw[fill=blue!20,rounded corners, ultra thick, dashed] 
			    (-0.5,0.5) rectangle (2.6,-0.5);
        \end{pgfonlayer}

\end{scope}
\begin{scope}[xshift=4cm]
		\node[vertex, label=above:a] (a) at (0,0) {};
		\node[vertex, black!50, label=left:v] (b) at (0.75,-1) {};
		\node[vertex, label=above:c] (c) at (1,0) {};
		\node[vertex, label=above:d] (d) at (1.5,0) {};
		\node[vertex, label=above:e] (e) at (2.2,0) {};
		
		\node[vertex, label=left:f] (f) at (0.5,-2) {};
		\node[terminal, label=right:g] (g) at (1.5,-2) {};
		
		\node[terminal, label=below:h] (h) at (1,-3) {};
		
		\node (txt) at (0,-3.5) {$\bar{G}_x$};

        \foreach \source/ \dest in {f/a, f/b, f/g, g/d,g/e,h/f,h/g}
            \path[edge, -] (\source) -- (\dest);
            
        \foreach \source/ \dest in {b/a, b/c, b/d}
            \path[edge, black!50, -] (\source) -- (\dest);
            
		\path[draw=blue!80, dashed, ultra thick, -] (0.2,0) -- (0.7,-0.7) -- (0.8,0);
		
		\path[draw=yellow!50, ultra thick, -] (1,-2.8) -- (0.7,-2) -- (0.95,-1);
		\path[draw=yellow!50, dashed, ultra thick, -] (0.95,-1) -- (1.2,0);
		\path[draw=yellow!50, ultra thick, -] (1.7,0) -- (1.58,-1.5) -- (2,0);
		
		\path[draw=green!50, ultra thick, -] (0.8,-3) -- (0.3,-2) -- (-0.2,0);
		
		\path[draw=red!80, ultra thick, -] (1.7,-2) -- (2.4,0);

        \begin{pgfonlayer}{background}
    		\path[fill=blue!20,rounded corners]
                (-0.5,-0.1) rectangle (2.6,-4);
			\draw[fill=blue!20,rounded corners, ultra thick, dashed] 
			    (-0.5,0.5) rectangle (2.6,-0.5);
        \end{pgfonlayer}
\end{scope}
\end{tikzpicture}
\caption{Illustration of a possible routing configuration extension during a forget node operation. The edges in grey are those belonging to~${E_v = \{va, vc, vd\}}$ and we have~${B_x = \{a, c,d,e\}}$,~${B_y = \{a, v, c,d,e\}}$ and~${d_{B_x}(v) = 3}$. The different routing paths are represented as colored lines. Dashed lines corresponds to a possible extension of these routing paths.}
\label{fig:algo-forget} 
\end{figure}

\paragraph{Join node} 
Let~$x$ be a join node with two children~$y, z$. By definition~$B_y = B_{z} = B_x$. For each~$\rc_y, \dc_y \in \rcset_y \times \dcset_y$ and each~$\rc_z, \dc_z \in \rcset_z \times \dcset_z$, let~$R_x = OPT_y(\rc_y, \dc_y) \cup OPT_z(\rc_z, \dc_z)$ and let~$\rc_x, \dc_y \in \rcset_x \times \dcset_x$ be the routing contract and delay contract induced by~$R_x$. We set~$OPT_x(\rc_x, \dc_x) = R_x$ if $R_x$ is valid and~${\mbox{\texttt{cong}}(R_x) < \mbox{\texttt{cong}}(OPT_x(\rc_x, \dc_x))}$.

%For all~$\rc \in \mathcal{P}(D) ^ {J_x^2}$ do the following. If~$A_y[\rc] = \emptyset$ then set~$A_x[\rc] = A_z[\rc]$ 
%otherwise if~$A_z[\rc] = \emptyset$ then set~$A_x[\rc] = A_z[\rc]$. 

%For all~$\rc \in \mathcal{P}(D) ^ {J_x^2}$, let~$R' = A_y[\rc] \cup A_z[\rc]$.
%If~$(\rc, R')$ is a valid record, then set~$A_x[\rc] = R'$.

\paragraph{Final step} 
Once we have computed the optimal solutions for every node, we can determine a complete routing configuration~$R^* \in \mathcal{R}(G,K)$ with minimum congestion for the instance~$I$ as follows.
Apply the forget node operation to every vertex in~$B_r$ to get a new table~$OPT$, then return the solution~$OPT(\rc, \dc)$ of minimum congestion among all~$\rc$ and~$\dc$. 
%is defined as~$\rc(\{s^k, t^k\}) = \{k\}$ for all~$k \in K$.

\textbf{Correctness.} %We prove that~$\rc_x$ and~$\dc_x$ are indeed the minimal information we need to solve each subproblem. Let~$x \in X$, a routing contract~$\rc_x \in \rcset_x$ is said to be valid if there exists a valid $\rc_x$-respecting routing configuration. The following claim is central to prove the correctness of the algorithm.
%The correctness of the algorithm follows from the fact that, at each step, we enumerate all the possible ways to extend previously computed routing configurations and the following claims.
The correctness follows from the following claim.

\begin{claim}
\label{claim:routig-pattern-feasible}
Let~$x \in X$ and~${R \in \mathcal{R} (\bar{G}_x, K)}$ be a minimum congested valid routing configuration. 
For every child~$y \in X$ of~$x$,~${OPT_y(\rc_y, \dc_y) \in \mathcal{R} (\bar{G}_y, K)}$ is a minimum congested valid routing configuration where~$\rc_y \in \rcset_y$ and~$\dc_y \in \dcset_y$ are induced by~${R\vert_{\bar{G}_y}}$.
\end{claim}
\begin{IEEEproof}
Let~$H$ and~$d$ be the routing contract and delay contract induced by~$R$.
Suppose that there exists a $\rc_y$-respecting and~$\dc_y$-respecting valid routing configuration~${R_y \in \mathcal{R} (\bar{G}_y, K)}$ such that~${\mbox{\texttt{cong}}(R_y) < \mbox{\texttt{cong}}(OPT_y(\rc_y, \dc_y))}$.
Consider the routing configuration~${R'\in \mathcal{R} (\bar{G}_x, K)}$ which is obtained by extending the routing paths of~$R_y$ the same way that~$OPT_y(\rc_y, \dc_y)$ gets extended to obtain~$R$. So~$R'$ is $H$-respecting,~$d$-respecting
%This is possible since~$R_y$ and~$OPT_y(\rc_y, \dc_y)$ both have the same structure (i.e they are both~$\rc_y$-respecting and~$\dc_y$-respecting).
and we have~${\mbox{\texttt{cong}}(R') < \mbox{\texttt{cong}}(R)}$.
We claim that~$R'$ is a valid routing configuration which contradicts the choice of~$R$ as being a minimum congested valid routing configuration in~${\mathcal{R} (\bar{G}_x, K)}$.
We show that~$R'$ is feasible since the other conditions of validity are satisfied by construction. Since~$R$ is feasible there exists a weight function~$w$ such that for each~${p^k \in R}$ the path~$p$ is the unique shortest path between its endpoints according to~$w$. We show how to construct a weight function~$w'$ from~$w$ so that~$R'$ is feasible with respect to~$w'$. For this purpose, we will use the fact that~$R$ and~$R'$ are both~$H$-respecting. For each~$uv \in E(H)$, there is corresponding routing subpath~$p[u, v]$ in~$R$ and~$p'[u, v]$ in~$R'$, and we set~$w'(e') = \frac{1}{\ell} \sum_{e \in p[u, v]} w'(e)$ for each edge~$e' \in p'[u, v]$ where~$\ell$ is the length of~$p[u, v]$. Finally, for every edge~$uv \in E(\bar{G}_x)$ such that~$w'(uv)$ is undefined, we set~$w'(uv) = +\infty$. This finishes the construction of~$w'$ and we claim that~$R'$ is feasible according to~$w'$. To see this, observe that any~$H$-respecting routing graph is obtainable from~$H$ by subdividing an appropriate number of time each edge in~$E(H)$. Thus, whenever there is a $H$-respecting routing graph that is feasible according to~$w$, it suffices to construct a weight function~$w'$ that preserves the distances between the vertices of degree greater than two.
The routing configuration~$R'$ is then valid and~${\mbox{\texttt{cong}}(R') < \mbox{\texttt{cong}}(R)}$ which contradicts the minimality of~$R$.
\end{IEEEproof}

%Suppose that there exists an $\rc_x$-respecting and $\dc_x$-respecting valid routing configuration~$R_x \in \mathcal{R}(\bar{G}_x, K)$ such that~$\mbox{\texttt{cong}}(R_x) < \mbox{\texttt{cong}}(OPT_x(\rc_x, \dc_x))$. Hence, 

%\begin{claim}
%\label{claim:dp}
%Let~$R_1, R_2$ be two feasible $\rc_x$-respecting and $\dc_x$-respecting routing configuration with the same congestion value. If $R_1$ can be extended to a complete routing configuration then so does $R_2$ with the same congestion value.
%\end{claim}
%\begin{IEEEproof}
%Let~$\bar{R}_1$ be a routing configuration such that~${R_1 \cup \bar{R}_1}$ is complete. Thus, by definition, there must exist a weight function $w:A \to \mathbb{Z}_{+}$ such that for each~${p^k \in R_1 \cup \bar{R}_1}$ the path~$p$ is the unique shortest path between its endpoints according to~$w$.
%We will show how to construct a new weight function~$w':A \to \mathbb{Z}_{+}$ such that for each~${p^k \in R_2 \cup \bar{R}_1}$ the path~$p$ is the unique shortest path between its endpoints according to~$w'$.

%\end{IEEEproof}

\textbf{Running time.} First, regarding the number of subproblems to solve, there are at most~$|\rcset_x| \cdot |\dcset_x|$ of them associated to each node~${x \in X}$. This corresponds to the number of possible pair routing contract and delay contract at each node~$x$. Since we have a nice tree decomposition of~$O(n)$ nodes, we end up with a total of at most~$O(|\rcset_x|\cdot |\dcset_x|\cdot n)$ subproblems to solve.
The most costly subproblem to solve is the forget node operation which takes time at most
$${(d_{B_x}(v) (d_{B_x}(v) - 1) / 2)^{|K_{closed}| + |K_{partial}|}} \cdot n^{O(1)}$$ 
which is bounded by $\omega^{O(|K_free|)} \cdot n^{O(1)}$.
Thus overall running time is
$$\omega^{O(|K_free|)} \cdot |\rcset_x|\cdot |\dcset_x|\cdot n^{O(1)}$$

%where we may have to update up to~$|\rcset_y|\cdot |\dcset_z|$ entries of table~$OPT_x$. Let~$b_x = |B_x| + |U_x|$, notice that~${b_x \leq \omega + |K_{free}|}$. We have,
%\begin{IEEEeqnarray}{rCl}
%|\rcset_y|\cdot |\rcset_z| &=& 2^{{b_y(b_y-1)/2}|K_{free}|} \cdot 2^{{b_z(b_z-1)/2}|K_{free}|}\\
%&=& 2^{{b_y(b_y-1)/2}|K_{free}| + {b_z(b_z-1)/2}|K_{free}|}\\
%&\leq&2^{{(\omega + |K_{free}|)(\omega + |K_{free}|-1)}|K_{free}|}\\
%&\leq&2^{{(\omega + |K_{free}|)^2}|K_{free}|}
%\end{IEEEeqnarray}

\begin{claim}
\label{claim:routig-contract-count}
Let~$x \in X$, we have $|\rcset_x| \leq 2^{O(|K||B_x|^8)}$
\end{claim}
\begin{IEEEproof}
By definition,~$\rcset_x$ contains only routing contracts that are induced by valid routing configurations in~${\mathcal{R} (\bar{G}_x, K)}$. Let~$H_x \in \rcset_x$ and~${R \in \mathcal{R} (\bar{G}_x, K)}$ be a valid~$H_x$-respecting routing configuration. First, the number of routings paths in~$R$ is bounded by~$O(|B_x|^2)$. Indeed, since~$R$ is valid, every~$p^k \in R$ is either complete or has at least one endpoint in~$B_x$. Hence, there can be as many routing subpaths as the number of pairs of vertices in~$B_x$ plus at most two routing paths per demand with exactly one endpoint in~$B_x$. Indeed, if there are more routing paths then we may create conflicting paths which is ruled out by Lemma~\ref{lem:no-conflict} or the graph induced by~$R$ may not contain only disjoint paths. Hence there can be no more than~$|B_x|(|B_x| - 1) / 2 + 2|B_x| = O(|B_x|^2)$ routing paths in~$R$ as claimed.
%The number of possible routing configurations in~$\mathcal{R} (\bar{G}_x, K)$ is then bounded by~$O(|B_x|^{2|K|})$. This corresponds to the number of possibilities to affect the demands to the routing paths.

Now we will determine the maximum number of possible routing contracts that can be obtained from valid routing configurations in~$\mathcal{R} (\bar{G}_x, K)$.
Let~$p$ be a path in~$H_x$ that is used to route some demand in~$K$.
By definition of a routing contract, we know that every vertex in~$p$ intersects with at least one other routing path in~$R$ (recall that all transit vertices are removed). 
Moreover, since there is no conflicting paths in~$R$ (Lemma~\ref{lem:no-conflict}), every other routing path can intersect~$p$ at most once. Thus~$p^k$ has no more than~$2|R| + 2$ vertices and then the graph~$H_x$ contains at most~$|R|(2|R| + 2) = O(|B_x|^4)$ vertices.
Finally, there can be at most~$2^{|K||E(H_x)|}$ possible edge-labelling function for~$H_x$. Hence, the number of possible routing contract in~$\rcset_x$ is bounded by~$2^{O(|K||B_x|^8)}$ as claimed.
\end{IEEEproof}

\begin{claim}
\label{claim:delay-contract-count}
Let~$x \in X$, we have $|\dcset_x| \leq |K| ^ {\Delta}$.
\end{claim}
\begin{IEEEproof}
By definition,~$\dcset_x$ contains only delay contracts that are induced by valid routing configurations in~${\mathcal{R} (\bar{G}_x, K)}$. Let~$\dc_x \in \dcset_x$ and~${R \in \mathcal{R} (\bar{G}_x, K)}$ be a valid~$\dc_x$-respecting routing configuration.
Thus, for all~${k \in K}$,  we have
$$d_x(k) = \sum_{p^k \in R} \sum_{uv \in p} \delta_{uv} \leqslant \Delta^{k} \leqslant \Delta$$ 
The number of possible delay contracts is then bounded by~$|K| ^ {\Delta}$ as claimed.
\end{IEEEproof}

%Furthermore, the tree decomposition has at most~$O(n)$ nodes and checking whether a routing configuration is feasible takes~$n^{O(1)}$ time, thus the overall running time is bounded by~\dynalgotimek, as claimed.

Using Claim~\ref{claim:routig-contract-count} and Claim~\ref{claim:delay-contract-count}, we deduce that the overall running time is bounded by~\dynalgotimek, as claimed.
\end{IEEEproof}

Since finding an optimal tree-decomposition of a graph is fixed-parameter tractable with respect to the treewidth of that graph \cite{bod1996}, we obtain the following result as an immediate corollary.

\begin{proposition} \label{th:fpt-tw}
The \kduspr problem is fixed-parameter tractable with respect to the parameter ``treewidth'' and ``number of demands''.
\end{proposition}

It is interesting to note that this algorithm can be compared with the two phases approach proposed in~\cite{Bley2011}. Indeed, the master problem that finds a set of routing paths is simply replaced here with a dynamic programming procedure while we still need the client to check for feasibility.

%% file: numerical_results.tex
In this section, we present some early experiments related to the D-USPR problem and based on the results described above. Both exact and heuristic solving approaches are implemented in Python and using Cplex 12.8 with the default settings and NetworkX graph library. We have tested our algorithms with the following features:
\begin{itemize}
    \item[$\bullet$] first by $(i)$ solving the basic formulation \eqref{ctn:obj}-\eqref{ctn:trivial_r},
    \item[$\bullet$] then by introducing $(ii)$ the subpath consistency inequalities \eqref{eq:subpath_1}-\eqref{eq:subpath_2}, $(iii)$ the node-precedence inequalities \eqref{eq:node_prec_1}-\eqref{eq:node_prec_2} and $(iv)$ both families of valid inequalities, to the basic formulation, 
    \item[$\bullet$] Algorithm \ref{alg:iterative_algorithm} uses a variant of the formulation \eqref{ctn:obj}-\eqref{ctn:trivial_r}, as described in Section \ref{sec:iterative_algorithm}, along with both families of valid inequalities.   
\end{itemize}
We have tested our algorithms on several instances derived from SNDlib\footnote{http://sndlib.zib.de} topologies of variying size and density. The big-M value is set to $|A|\times |K|$ for all the experiments, likewise in \cite{Bley2011} and the CPU time limit is fixed to 5 hours for the exact approach. 

\newcolumntype{d}[1]{D{.}{\cdot}{#1}}
%\newcolumntype{d}[1]{D{.}{.}{#1}}
{\setlength{\extrarowheight}{1.5pt}
\begin{table*}[htbp]
  \centering
   \caption{\small{The impact of each class of valid inequalities}}
  \scriptsize{
    \begin{tabular}{l*{13}{r}} %@{$\Longrightarrow\,\,\,$} {lld{5}d{3}d{5}d{3}}
    \toprule
     \multicolumn{4}{l}{} & \multicolumn{3}{c}{$(i)$ Basic MILP} &       \multicolumn{3}{c}{$(ii)$ MILP + subpath cons. ineq.} &       \multicolumn{3}{c}{$(iii)$ MILP + node precendence ineq.} \\
        \cmidrule(r){5-7} \cmidrule(r){8-10} \cmidrule(r){11-13}
    \multicolumn{1}{l}{Topology}	&	\multicolumn{1}{c}{$|V|$}	&	\multicolumn{1}{c}{$|A|$}	&	\multicolumn{1}{c}{$|K|$}	& \multicolumn{1}{c}{Gap (\%)}	&	\multicolumn{1}{c}{Nodes}	&	\multicolumn{1}{c}{TT} &	\multicolumn{1}{c}{Gap (\%)}	&	\multicolumn{1}{c}{Nodes}	&	\multicolumn{1}{c}{TT} & \multicolumn{1}{c}{Gap (\%)}	&	\multicolumn{1}{c}{Nodes}	&	\multicolumn{1}{c}{TT}	\\
\midrule													
PDH	&	11	&	68	&	24	&	0.00	&	1	&	0.80	&	0.00	&	1	&	0.41	&	0.00	&	1	&	0.34	\\
Di-yuan	&	11	&	84	&	22	&	0.00	&	6571	&	773.41	&	0.00	&	1	&	1.24	&	0.00	&	1	&	0.45	\\
Polska	&	12	&	36	&	66	&	6.70	&	602765	&	17302.36	&	6.62	&	71	&	34.66	&	6.62	&	78295	&	 6780.74	\\
Nobel-US	&	14	&	42	&	91	&	7.98	&	19004	&	18000.00	&	2.02	&	5	&	439.08	&	-	&	255813	&	18000.00	\\
Dfn-bwin	&	10	&	90	&	90	&	0.00	&	1	&	0.45	&	51.25	&	1	&	18000.00	&	0.00	&	1	&	1.91	\\
abilene	&	12	&	30	&	132	&	0.00	&	6832	&	19.49	&	0.00	&	1	&	8.59	&	0.00	&	6271	&	17.77	\\
Dfn-gwin	&	11	&	94	&	110	&	6.67	&	5375	&	862.72	&	37.82	&	1	&	18000.00	&	12.03	&	31	&	707.97	\\
Atlanta	&	15	&	44	&	210	&	0.00	&	10839	&	187.22	&	0.25	&	1	&	20.58	&	0.00	&	125393	&	6280.28	\\
Nobel-GER	&	17	&	52	&	121	&	-	&	-	&	18000.00	&	12.12	&	1	&	811.39	&	-	&	385146	&	18000.00	\\
    \bottomrule
    \end{tabular}%
    %\vspace*{0.3cm}
     
    }
  \label{tab:efficiency_per_class}%
\end{table*}%
}
Table \ref{tab:efficiency_per_class} shows the impact of using valid inequalities and the efficiency of each class in strengthening the basic formulation \eqref{ctn:obj}-\eqref{ctn:trivial_r} for nine instances. The first four columns refer to the name, number of nodes, number of arcs and number commodities, for each instance. Then, for each of the configurations $(i)$ (basic MILP), $(ii)$ (MILP + subpath consistency inequalities) and $(iii)$ (MILP + node precedence inequalities), we show the following entries: Gap (\%) is the root gap (the relative difference between the best upper bound (optimal solution if the problem has been solved to optimality) and the lower bound obtained at the root node), Nodes is the number of nodes in the branch-and-bound tree and TT is the CPU time for computation (in seconds). The value in Gap column is written in italics if the solution found within the time limit was not optimal and replaced by ``--'' if no feasible solution was found within that time. We can see from Table \ref{tab:efficiency_per_class} that the subpath consistency constraints \eqref{eq:subpath_1}-\eqref{eq:subpath_2} allow to improve the gap for several instances and help in reducing substantially both the number of nodes in the branch-and-bound tree and the CPU time for computation. In particular, except for \textit{dfn-bwin} and \textit{dfn-gwin}, that are the instances with highest density, all the instances are solved to optimality in less than 20 minutes, most of them at the root node. A less significant impact is obtained when using node precedence inequalities \eqref{eq:node_prec_1}-\eqref{eq:node_prec_2}, yet they allow to speed up the resolution for several instances, and to reduce the size of the branch-and-bound tree (like for instance \textit{di-yuan}). 

Table \ref{tab:tab2} shows the results obtained when adding both families of inequalities to the basic formulation for the previous instances. We can see from the table that there is a positive but slight impact on the CPU time, especially for instances \textit{Di-yuan} and \textit{Nobel-US}.   
\begin{table}[htbp]%htbp
		\caption{The impact of adding both classes of inequalities}  
		\begin{center}
		  \scriptsize{
		%	\resizebox{.8\textwidth}{!}{
				%\resizebox{\linewidth}{!}{
				\begin{tabular}[t]{l|*{5}{c}}
					\toprule
					Name & LB & UB	& root gap (\%) & Nodes & TT (sec.) \\	
					\midrule									
PDH	& 12.80	&	12.80	&	0.00	&	1	&	0.50 \\
Di-yuan	& 5.00	&	5.00	&	0.00	&	1	&	0.30	\\
Polska	& 6.41	&	6.90	&	6.70	&	243	&	60.97 \\
Nobel-US	& 24.20	&	24.70	&	2.02	&	5	&	263.86 \\
Dfn-bwin	& 0.34	&	0.69	&	50.72	&	1	&	18000.00 \\
abilene	&	60.41	&	60.4	&	0	&	1	&	10.17	\\
Dfn-gwin & 0.65	&	1.05	&	38.10	&	1	&	18000.00 \\
Atlanta	&	3.58	&	3.58	&	0,00	&	5	&	29.64 \\
Nobel-GER	&	3.87	&	4.40	&	12.12	&	16	&	458.00 \\
					\bottomrule
				\end{tabular}
			}
		\end{center}
		\label{tab:tab2}
	\end{table}
	
Note that, these results although promising can be significantly improved by generating the valid inequalities in a dynamic fashion (via separation routines in a branch-and-cut framework) instead of being all integrated in the basic MILP.

We have tested our iterative algorithm on instances \textit{France}, \textit{Nobel-EU} and \textit{Norway} that could not be solved using the exact approach due to their size, density and number of demands. Those three instances are among the most challenging state-of-the-art instances for the USPR problem. The iterative algorithm allows to obtain a good upper bound for all three instances simply by improving an existing solution (e.g. one obtained from the minimum spanning tree congestion). A preliminary set of results shows empirically that a good partition of the demands set with an appropriate selection of the demands to reroute ($K_{free}$) allows to substantially improve the trivial bound of the existing solution. In addition, the fact that we start from an existing solution allows us to save computation efforts highlighting even further the potential scalability of our algorithm. For example, in instance France, rerouting 3$\%$ of the demands allows to improve the minimum spanning tree congestion bound by 2$\%$ while an improvment of 7$\%$ is enabled by rerouting 16$\%$ of the demands. 

%% file: conclusion.tex
In this paper, we have investigated several research directions to go towards more scalable algorithmic solutions. For this purpose, we proposed the following two approches: $(i)$ reducing the size of the problem (e.g number of demands) and $(ii)$ exploiting the structure of the input graph (e.g treewidth). Although the obtained results are promising there is still room for improvment.
First, we expect that solving the formulation using a branch-and-cut algorithm will subtantially improve the performance of the MILP-based exact approach and the efficiency of the iterative algorithm.
Second, we are implementing the dynamic programming algorithm and even though the theoretical complexity is prohibitive, we believe that the efficiency of this algorithm can be good in practice especially if used in combination with a parallelization approach.
Finally, this problem seems to be a good candidate to apply machine learning methods in the hope to reach better running time through learned heuristics.

%% file: conference_101719.bbl
\begin{thebibliography}{10}

\bibitem{benameur03}
W.~Ben{-}Ameur and {\'{E}}.~Gourdin.
\newblock Internet routing and related topology issues.
\newblock {\em {SIAM} J. Discrete Math.}, 17(1):18--49, 2003.

\bibitem{BleyPhdthesis}
A.~Bley.
\newblock {\em Routing and capacity optimization for IP networks}.
\newblock PhD thesis, Technische Univertität Berlin, 2007.

\bibitem{bley2009}
A.~Bley.
\newblock Approximability of unsplittable shortest path routing problems.
\newblock {\em Networks}, 54(1):23--46, 2009.

\bibitem{Bley2011}
A.~Bley.
\newblock An integer programming algorithm for routing optimization in {IP}
  networks.
\newblock {\em Algorithmica}, 60(1):21--45, 2011.

\bibitem{Bley2010}
A.~Bley, B.~Fortz, E.~Gourdin, K.~Holmberg, O.~Klopfenstein, M.~Pi{\'o}ro,
  A.~Tomaszewski, and H.~{\"U}mit.
\newblock {\em Optimization of OSPF Routing in IP Networks}, pages 199--240.
\newblock Springer Berlin Heidelberg, Berlin, Heidelberg, 2010.

\bibitem{Bley1998}
A.~Bley, M.~Gr{\"o}tschel, and R.~Wess{\"a}ly.
\newblock Design of broadband virtual private networks: Model and heuristics
  for the b-win.
\newblock In {\em Robust Communication Networks: Interconnection and
  Survivability}, 1998.

\bibitem{bod1996}
H.~L. Bodlaender.
\newblock A linear-time algorithm for finding tree-decompositions of small
  treewidth.
\newblock {\em {SIAM} Journal on Computing}, 25(6):1305--1317, 1996.

\bibitem{DowneyF13}
R.~G. Downey and M.~R. Fellows.
\newblock {\em Fundamentals of Parameterized Complexity}.
\newblock Springer-Verlag, 2013.

\bibitem{Fortz2000}
B.~Fortz and M.~Thorup.
\newblock Internet traffic engineering by optimizing {OSPF} weights.
\newblock In {\em Proceedings {IEEE} {INFOCOM} 2000, The Conference on Computer
  Communications}, pages 519--528. {IEEE} Computer Society, 2000.

\bibitem{Garcia2009}
R.~Garcia.
\newblock {\em Resource constrained shortest paths and extensions}.
\newblock PhD thesis, Georgia Institute of Technology, Atlanta, GA, {USA},
  2009.

\bibitem{Horvath2016}
M.~Horváth and T.~Kis.
\newblock Solving resource constrained shortest path problems with lp-based
  methods.
\newblock {\em Computers \& Operations Research}, 73:150 -- 164, 2016.

\bibitem{kloks94}
T.~Kloks.
\newblock {\em {Treewidth, Computations and Approximations}}.
\newblock Springer, 1994.

\bibitem{Parmar2006}
A.~Parmar, S.~Ahmed, and J.~Sokol.
\newblock An integer programming approach to the ospf weight setting problem.
\newblock 2006.

\bibitem{Perrot2019}
N.~Perrot, A.~Benhamiche, Y.~Carlinet, and E.~Gourdin.
\newblock {\em Future Networks: Overview of Optimization Problems in
  Decision-Making Procedures}, pages 177--207.
\newblock IGI Global, 2019.

\end{thebibliography}
